\pdfoutput=1

\documentclass[11pt]{article}

\usepackage[final]{acl}

\usepackage{times}
\usepackage{latexsym}
\usepackage{amsmath} 
\usepackage{amssymb}
\usepackage[ruled,vlined]{algorithm2e}
\usepackage{multirow}
\usepackage{booktabs,caption}
\usepackage[flushleft]{threeparttable}

\usepackage[T1]{fontenc}

\usepackage[utf8]{inputenc}

\usepackage{microtype}

\usepackage{inconsolata}

\usepackage{graphicx}

\title{Visual Zero-Shot E-Commerce Product Attribute Value Extraction}

\author{\bf Jiaying Gong$^{1}$, Ming Cheng$^{2}$, Hongda Shen$^{1}$\\ \bf Pierre-Yves Vandenbussche$^{1}$,  Janet Jenq$^{1}$, Hoda Eldardiry$^{2}$ \\
$^{1}$eBay Inc., $^{2}$Virginia Tech \\
\texttt{\{jiagong,honshen,pvandenbussche,jjenq\}@ebay.com}, \\
\texttt{\{ming98,hdardiry\}@vt.edu}
}

\begin{document}
\maketitle
\begin{abstract}
Existing zero-shot product attribute value (aspect) extraction approaches in e-Commerce industry
rely on uni-modal or multi-modal models, where the sellers are asked to provide detailed textual inputs (product descriptions) for the products.
However, manually providing (typing) the product descriptions is time-consuming and frustrating for the sellers.
Thus, we propose a cross-modal zero-shot attribute value generation framework (ViOC-AG) based on CLIP, which only requires product images as the inputs.
ViOC-AG follows a text-only training process, where a task-customized text decoder is trained with the frozen CLIP text encoder to alleviate the modality gap and task disconnection.
During the zero-shot inference, product aspects are generated by the frozen CLIP image encoder connected with the trained task-customized text decoder.
OCR tokens and outputs from a frozen prompt-based LLM correct the decoded outputs for out-of-domain attribute values.
Experiments show that ViOC-AG significantly outperforms other fine-tuned vision-language models for zero-shot attribute value extraction.
\end{abstract}

\section{Introduction}
Product attribute value extraction aims at retrieving the values of attributes from the product's unstructured information (e.g. title, description), to serve better product search and recommendations for buyers.
Existing uni-modal or multi-modal attribute value extraction models require sellers to manually provide (type) product descriptions, which is time-consuming and frustrating.
In addition, these approaches mainly focus on supervised learning, weakly-supervised learning, and few-shot learning to train or fine-tune language models for attribute value prediction~\cite{yang2023mixpave, 10.1145/3583780.3615142, xu2023towards}.
These approaches need labeled data for training and can not be extended to unseen attribute values for new products.
To extract unseen attribute values, text-mining models~\cite{li2023attgen, xu2023towards}, inductive graph-based models~\cite{hu2025hypergraphbased, gong2024multi}, and multi-modal large language models~\cite{zou2024implicitave, zou2024eiven} try to generate potential attribute values from both product descriptions and images.

\begin{figure}[htp] 
 \center{\includegraphics[height=3.5cm,width=7.7cm]{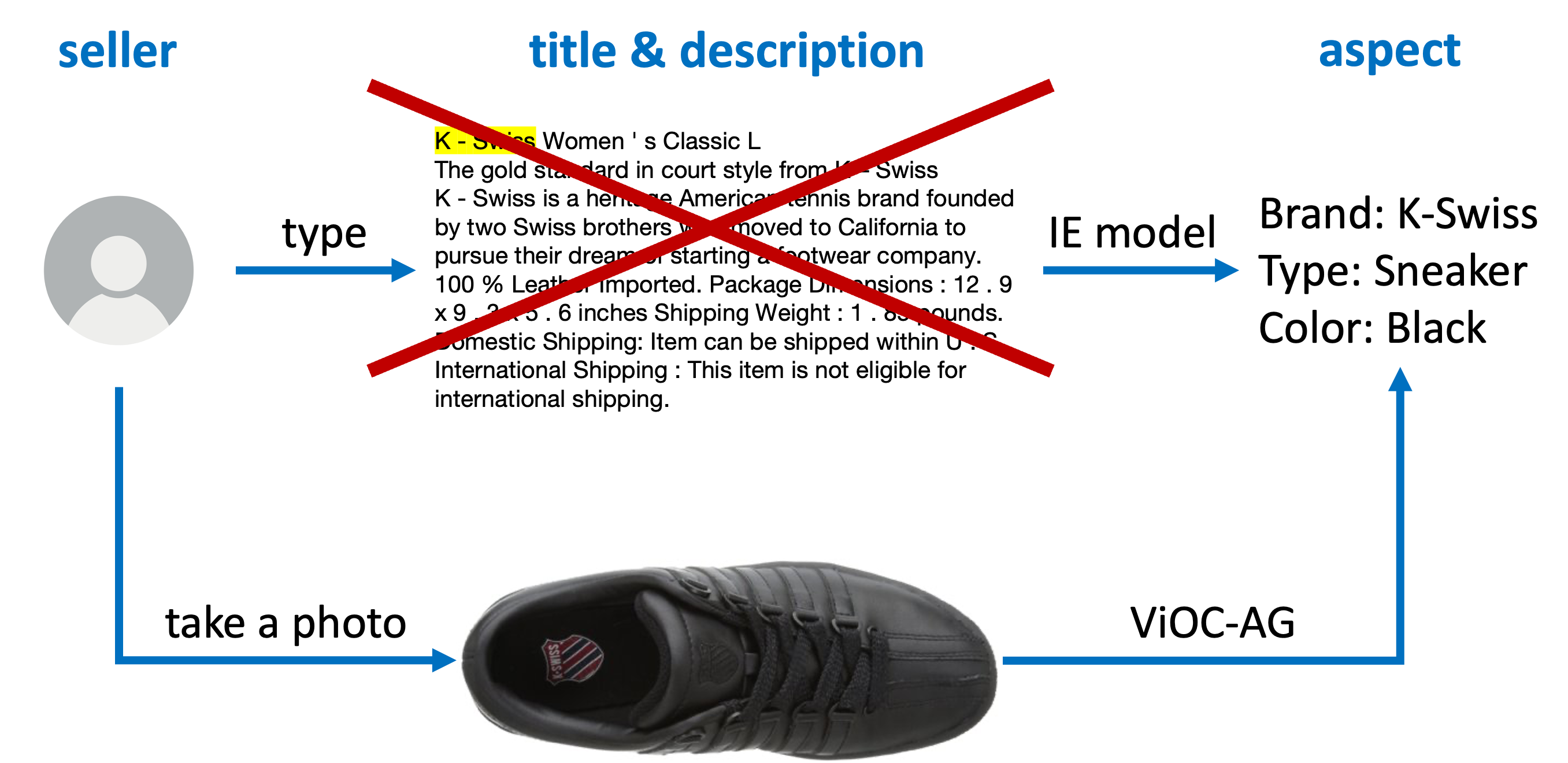}}
 \caption{\label{fig:example} An example of cross-modal aspect generation.}
 \vspace{-4mm}
 \end{figure}

However, these approaches suffer from several limitations: (1) it is difficult for classification or graph-based prediction models to scale to a large number of attribute values because the decision boundaries between classes become more complex and harder to learn, and increase the computational complexity.
(2) traditional information extraction models or the above multi-modal models need the inputs for product textual descriptions from the sellers (see Figure~\ref{fig:example}).
It is challenging and time-consuming for the sellers to manually type and provide the product descriptions because sometimes sellers themselves don't know the correct answers, which may cause ambiguity for attribute values.
To address the above limitations, we propose an OCR and product captions enhanced zero-shot cross-modal model (ViOC-AG) to generate attribute values, which ONLY need the product images as the inputs. 
In other words, the seller only needs to take a photo of the product that he wants to sell without manually providing the product textual descriptions, resulting in a better user experience.

There are two main challenges for zero-shot cross-modal aspect generation. 
The first challenge is the modality gap between vision and language caused by cross-modal generation.
Although there exist many large generative image-to-text transformers (i.e. BLIP-2~\cite{li2023blip}), they target at the image captioning or visual question answering tasks.
Our experiments in Sec.~\ref{sec:experiments} show that simply fine-tuning these large vision language models performs poorly on the product attribute value generation task.
This is because there is a task disconnection between language modeling (used for image captioning) and aspect generation.
Thus, we take advantage of the pre-trained CLIP~\cite{radford2021learning} ability to align visual and textual representations in a shared embedding space to avoid the modality gap.
To alleviate task disconnection, we train a task-customized text decoder with a projection layer, which follows a text-only training process.
Specifically, we tend to transfer CLIP textual description embeddings back into textual aspects by learning a task-customized decoder for the frozen CLIP text encoder using only text.

The second challenge is the out-of-domain aspects caused by zero-shot generation.
For zero-shot aspects, the model is susceptible to generate aspects that are not actually present in the input image but frequently appear during training (object hallucination).
Due to the characteristics of the product attribute value generation task, some aspects (i.e. brand, capacity, etc.) are shown directly on the product.
Thus, we correct the generated outputs from the trained task-customized text decoder with the OCR tokens. 
For further final aspects correction, we generate potential attribute value answers by designing prompt templates for pre-trained visual question-answering LLMs.
The effectiveness of each module is shown independently in Sec.~\ref{sec:results}.
Extensive experimental results on  MAVE~\cite{yang2022mave} dataset show that our proposed model ViOC-AG significantly outperforms other existing vision language models for zero-shot attribute value generation. 
ViOC-AG also achieves competitive results with generative LLMs with textual product description inputs, showing the positive potential that users only need to take photos of the selling products for aspect generation.

\section{Related Works}
Existing works on product attribute value extraction mainly focus on supervised learning to train classification models~\cite{10386204, chen2022extreme, 10020304}, QA-based models~\cite{chen-etal-2023-named, liu2023knowledge, shinzato2022simple, wang2020learning} or large language models~\cite{fang2024llm, brinkmann2023product, baumann2024using}. 
However, these approaches require large quantities of labeled data for training.
Recently, some works use few-shot learning~\cite{10.1145/3583780.3615142, yang2023mixpave} and weakly supervised learning~\cite{xu2023towards, zhang2022oa} to reduce the amount of labeled data for training. 
But these approaches still need labeled data for multi-task training or iterative training.

To extract unseen attribute values, text-mining models~\cite{li2023attgen, xu2023towards} extract explicit attribute values directly from text, and zero-shot models~\cite{hu2025hypergraphbased, gong2024multi} predict new attribute values by inductive link prediction of graphs.
However, all these approaches can only extract attribute values from textual inputs. 
In other words, these models are from a single modality. 
Then, some multi-modal models use both the product image and title with the description as the inputs to learn a better product representation for attribute value extraction~\cite{zou2024implicitave, zou2024eiven, liu2023multimodal, wang2023mpkgac, ghosh2023d, wang2022smartave, liu2022boosting}.
Though performance is improved by fusing more semantic information from multiple modalities, more input data is needed during the training stage. 
To enable image-first interactions from sellers and make it simple for the users, we propose a zero-shot cross-modal model motivated by image captioning~\cite{fei2023transferable, guo2023images, xu2023zero, zeng2023conzic, tewel2022zerocap} for attribute value generation, where only images are used as inputs.

\section{Methodology}
\begin{figure*}[htp] 
 \center{\includegraphics[height=6.2cm,width=\textwidth]{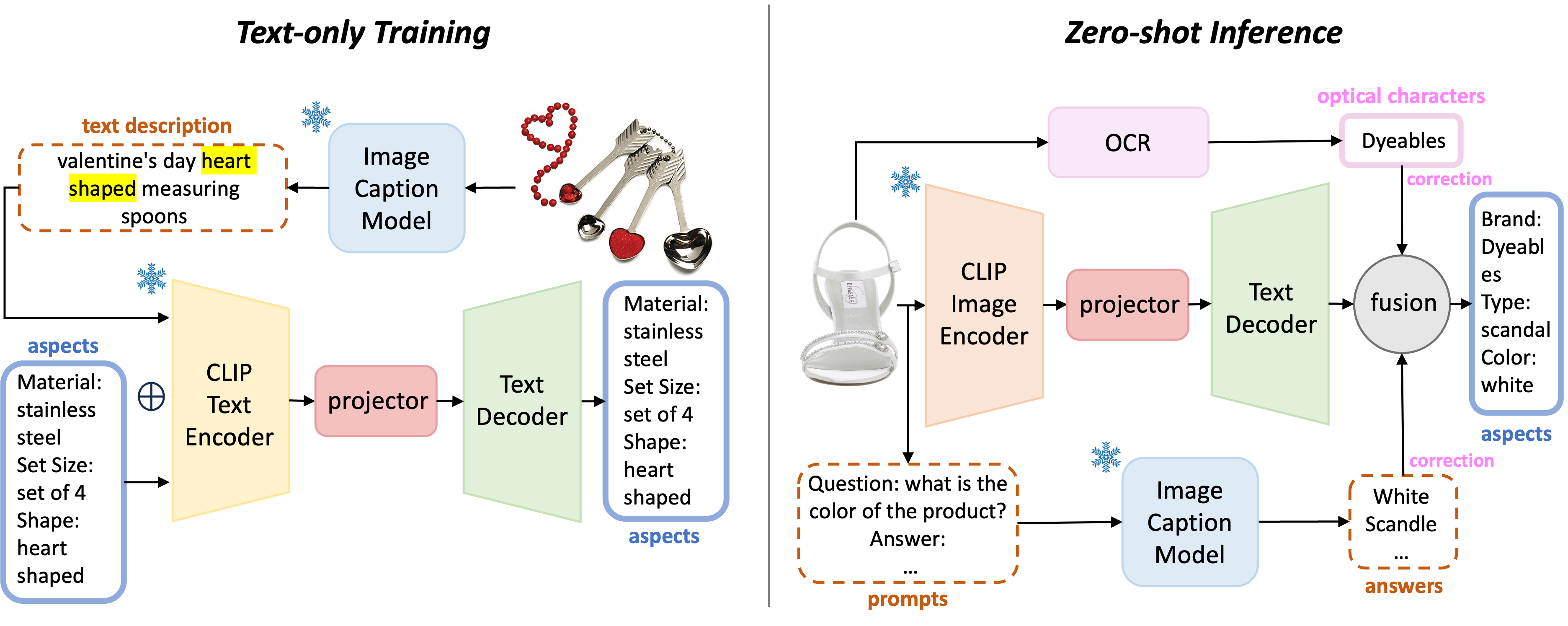}}
 \caption{\label{fig:overall} The overview of our proposed ViOC-AG model. Only the projector and the text decoder are trainable.}
 \end{figure*}
\subsection{Problem Definition}
Cross-modal attribute-value generation aims at automatically generating textual product attribute values from the product image.
Consider a dataset $\mathcal{D} \subset \mathcal{I} \times \mathcal{T}$ where $\mathcal{I}$ is the image domain and $\mathcal{T}$ is the text domain, and $(I_{i}, A_{i})$ forms a corresponding image-aspect pair (i.e. $A_{i} \in \mathcal{T}$ is attribute values from product $I_{i}$).
It can be formalized as a sequence generation problem given an input image $I$ with a set of detected OCR tokens $T$, the model needs to infer the attribute values $A=[a_1, \cdots, a_N]$, where $a_N$ denotes each attribute-value and $N$ is the number of attribute-values.
The problem focuses on searching $A$ by maximizing $p(A|I)$:
\begin{equation}
\text{log}p(A|I)=\text{log}\prod_{N}^{}p(a_{N}|I,T,a_{1:n-1})
\end{equation}
where $T$ is the set of OCR tokens detected from the product image $I$. 
The training process is typically accomplished in a supervised manner by training on manually annotated datasets and optimizing weights to converge to the optimal state.
Therefore, it is necessary to explore optical-characters-aware zero-shot methods for guiding large-scale language models free of parameter optimization.

\subsection{Zero-Shot Data Sampling and Pre-processing}~\label{sec:data_sampling}
For zero-shot attribute-value(aspect) generation, we follow~\cite{gong2024multi} to let $A^{S} = [a^{S}_{1}, \cdots, a^{S}_{N}]$ and $A^{U} = [a^{U}_{1}, \cdots, a^{U}_{N}]$ denote the seen aspects and unseen aspects, where $A^{S} \cap A^{U} = \varnothing$. 
Because one product may contain multiple aspects, We follow a generalized zero-shot setting~\cite{pourpanah2022review} to ensure that any product in the validation/testing set has at least one aspect from $A^{U}$.
For data pre-processing, we first combine the aspects that only have differences in uppercase/lowercase, singular/plural forms, or similar meanings and drop the data that we can not retrieve the corresponding images by the provided URLs in MAVE~\cite{yang2022mave}. 
We implement the zero-shot data sampling over 21 categories of MAVE independently so that the zero-shot training, validation, and testing sets can still have similar data distributions across various categories.

\subsection{Overall Framework}
We introduce the overview of ViOC-AG in Figure~\ref{fig:overall}, which is a transferable aspect generation framework based on CLIP~\cite{radford2021learning} and trained on a text-only corpus.
Both encoders in CLIP are trained jointly using a contrastive loss to ensure that the representations of an image and its corresponding text are close in the feature space.
We train a language decoder to decode the CLIP text embedding of aspects with generated text descriptions from a frozen image caption model.
We make this decoding to be similar to the original textual aspects $A$. 
Namely, our training objective is a reconstruction of the input text from CLIP textual embedding.
For zero-shot inference, we directly feed the CLIP image embedding of a given product image $I$ into the trained decoder to generate aspects that are corrected by detected optical characters and values from the generated text description.

\subsubsection{Text-only Training}
Our goal is to train a transferable task-customized language decoder with a projector.
During the training phase, we freeze all the parameters of the CLIP text encoder.
We only train the projector from scratch and fine-tune the decoder-only language model (i.e. GPT-2) in predicting product attribute values.
We first concatenate the generated descriptions of the product image via a frozen image caption model with the textual aspects inputs sequentially to prevent model overfitting and improve the generalization and robustness of the model. 
Next, we mapped the textual embeddings to CLIP space by CLIP text encoder $E_{T}^{*}$.
A projection layer is also trained for dimension alignment and alleviating the modality gap.
Then, the projected text embedding is decoded back by a trainable decoder $D_{T}$.
The text-only training objective is thus to minimize:
\begin{equation}
\sum_{A \in \mathcal{T}}^{}\mathcal{L}(D_{T}(W \cdot E_{T}^{ *}(A\oplus M^{* }(I))+b),A)
\end{equation}
where $*$ denotes a frozen model with parameters not updated during training. $M^{*}$ can be any frozen image caption model (i.e. BLIP-2), and $I$ is the product image.
The projector $W(\cdot) + b$ is a learnable linear layer for domain alignment and dimension adjustment.
$\mathcal{L}$ is an autoregressive cross-entropy loss for all tokens in $A$.

\subsubsection{Zero-shot Inference}~\label{sec:zero-shot}
After the decoder $D_{T}$ is trained, we can leverage it for zero-shot generation inference.
Given a product image $I$, we first extract its visual embeddings via the frozen CLIP image encoder $E_{I}^{*}$.
We then employ the trained projector and text decoder $D_{T}$ to convert the visual embeddings into textual aspects:
\begin{equation}
A_{D} = D_{T}(W \cdot E_{I}^{*}(I)+b)
\end{equation}
where $W(\cdot) + b$ is the trained projector. To improve the zero-shot performance caused by the out-of-domain attribute values, a fusion module is employed to correct the outputs from the text decoder $D_{T}$.
We use information from two major sources to correct the outputs from $A_{D}$ for the final aspects:
(1) the values generated by the frozen prompt-instructed image caption model $A_{P} = \text{LLM}(I, P)$, where LLM can be any frozen cross-modal model (i.e. BLIP-2, LLaVA, etc.)~\footnote{We use BLIP-2 as the image caption model in our paper.}, and $P$ are the prompt templates (i.e. Question: What is the \textit{attribute} of the product? Answer:). 
The \textit{attribute} is replaced with the collected attribute names (i.e. type, brand, color, etc.) in the training set;
(2) the optical characters $T$ detected by the OCR module:~\footnote{https://github.com/JaidedAI/EasyOCR}
\begin{equation}
T = \text{OCR}(I) = \left \{ t | c_{t} > \tau_{c} \right \}
\end{equation}
where $c_{t}$ is token confidence value, and $\tau_{c}$ is the confidence threshold.
\begin{algorithm}[!htbp]
\small
\SetKwInOut{KIN}{Input}
\SetKwInOut{KOUT}{Output}
\caption{Zero-shot Inference Correction}
\label{alg:zeroshot} 

\KIN{Aspects $A_{D}$, $A_{P}$, OCR tokens $T$ and distance threshold $\tau_{d}$}
\KOUT{Final Aspects $A$}

\For{$a_D$ in $A_D$}
{
   
        \eIf {get\_attribute($a_{D}$) $\in$ get\_attribute($A_{P}$)}
        {
            \eIf{cosine\_similarity(get\_value($a_{D}$), get\_value($a_{P}$)) $ >  \tau_{d}$}
            {
            A.update($a_{P}$)
            }
            {
            A.update($a_{i} | max(\text{cosine\_similarity}(a_D$, $a_P || T$)))
            }
            
        }
        {A.update($a_{i} | max(\text{cosine\_similarity}(a_D$, $T$)))}
        {

        }
        
}
return $A$
\end{algorithm}

In most cases, product attributes are from a known set (i.e. type, brand, etc.), only the values (i.e. long wallet, Chanel, etc.) vary for different products and may include zero-shot cases, such as a new brand.
We first check whether the attribute exists in the training set to decide whether the attribute is a zero-shot case or not.
When the attribute is not a zero-shot case, we further compare the cosine similarity between $A_{D}$ and $A_{P}$. If the value is closer to 1, $A_{P}$ is used to correct $A_{D}$ for irrelevant tokens.
If they are quite different, we consider it as a zero-shot case, where OCR tokens $T$ are used to further correct $A_{D}$.
For attribute value zero-shot cases, only OCR tokens $T$ are used to correct $A_{D}$ because no relevant prompts are provided for the generated $A_{P}$.
Details of the correction is shown in Algorithm~\ref{alg:zeroshot}.
The correction process solves the hallucination problem and improves the zero-shot performance on out-of-domain attribute values. 

\section{Experiments}~\label{sec:experiments}
\vspace{-5mm}
\subsection{Experimental Setup}
\subsubsection{Dataset}
We evaluate our model over MAVE, which is a multi-label large e-commerce dataset derived from Amazon Review Dataset~\cite{yang2022mave}. 
To simulate the zero-shot situation, we reconstruct the dataset into zero-shot learning settings followed by Sec.~\ref{sec:data_sampling}, where
there is no overlap of classes between the training and the testing set.
Dataset statistics and label counts distributions are shown in Sec.~\ref{sec:data} in Appendix.

\subsubsection{Baselines and Evaluation Metrics}
We compare our model ViOC-AG with the following open-sourced generative vision language models: ViT-GPT~\cite{dosovitskiy2020image, radford2019language}, GIT~\cite{wang2022git}, LLaVA~\cite{liu2024visual}, BLIP~\cite{li2022blip}, BLIP-2~\cite{li2023blip}, and InstructBLIP~\cite{dai2024instructblip}.
We additional compare ViOC-AG with some text-based LLMs (BART~\cite{lewis2019bart} and T5~\cite{raffel2020exploring}), which use product titles as the inputs, to explore whether only using visual inputs can achieve competitive results.



For evaluation, we use 80\% Accuracy (we assume it is correct when 80\% of the generated outputs are matched with the golden label for one aspect) to measure the generation accuracy.
Besides, we use Micro F1 and Macro F1 to evaluate the retrieval performance.
We also use ROUGE1~\cite{lin-2004-rouge} to evaluate the generation quality.
We provide explanations in Sec.~\ref{sec:evaluation} in Appendix.
Parameter settings are provided in Sec.~\ref{sec:parameter} in Appendix. 
For deploying ViOC-AG at scale, The pre-trained image caption model needs at least V100 GPUs are needed for inference. No GPU is required for the OCR module. A100 or V100 GPUs are needed for the textual decoder training. 

\subsection{Results and Discussions}~\label{sec:results}
\vspace{-5mm}
\subsubsection{Main Results}~\label{sec:main}
\begin{table}[]
\small
\caption{Experimental results (\%) of text-only models and image-to-text models on the MAVE dataset.}
\label{tab:main}
\centering
\tabcolsep=0.08cm
\begin{tabular}{lcccc}
\hline
               & 80\%Acc.       & Macro-F1       & Micro-F1       & ROUGE1         \\ \hline
BART           & \textbf{79.32} & 13.24          & 19.54          & \textbf{60.59} \\
T5             & 68.69          & \textbf{15.28} & \textbf{23.06} & 53.82          \\ \hline
ViT-GPT        &    16.60            & 2.62           & 4.07           & 31.00          \\
GIT            & 14.89          & 3.70           & 5.36           & 34.13          \\
LLaVA          &     25.67           &   7.20             &     10.24           &      40.11          \\
BLIP           & 33.13          & 8.92           & 12.42          & 38.56          \\
InstructBLIP   &   40.00             &   12.54             &    17.05            &     \textbf{44.20}           \\
BLIP-2         & 45.85          & 13.92          & 18.86          & 43.06          \\
ViOC-AG (ours) & \textbf{54.82} & \textbf{17.71} & \textbf{23.69} & 31.92          \\ \hline
\end{tabular}
\end{table}

The results of zero-shot attribute value prediction are shown in Table~\ref{tab:main}.
We observe that:

(1) In general, text-only models (BART and T5) show better performance than image-to-text models. This is because there is no modality gap for text-only models as they sacrifice the user experience that product text descriptions are needed for the model inputs.
Thus, our goal is to build an image-to-text (cross-modal) model requiring only image inputs (product photos), which can achieve at least a similar performance to text-only models.

(2) Although existing vision-language models (i.e. BLIP, LLaVa) have the zero-shot ability in image captioning, they perform poorly on product attribute value generation. 
We think that this is because there is a task disconnection between the image captioning task and the attribute value generation task.
Simply fine-tuning the vision language models may improve the image caption task.
However, task-oriented information (i.e. OCR from the product, task-customized decoder, etc.) is also important for product attribute value generation tasks.

(3) Our proposed model achieves the best Macro and Micro F1 scores among all text-only and image-to-text models, but it has a lower accuracy and ROUGE value compared with text-only models.
We conjecture that this is because the trained task-customized text decoder may generate some non-relevant tokens, which reduces the percentage of the accurate tokens among all generated outputs, resulting in a low ROUGE and accuracy.
More effective post-processing techniques can be studied in future work to remove the non-relevant tokens.

\begin{table}[]
\caption{Performance metrics (\%) of the proposed approach over ten categories on MAVE dataset.}
\small
\label{tab:category}
\centering
\tabcolsep=0.08cm
\begin{tabular}{lcccc}
\hline
             & 80\%Acc.  & Macro-F1 & Micro-F1 & ROUGE \\ \hline
Industrial   & 34.51 & 10.64  & 15.12  & 24.65 \\
Home Kitchen & 42.25 & 11.76  & 16.19  & 23.56 \\
Automotive   & 43.64 & 13.28  & 17.49  & 28.81 \\
Musical      & 51.74 & 14.65  & 20.08  & 30.76 \\
Sports       & 47.38 & 16.08  & 21.73  & 30.16 \\ \hline
Pet          & 64.45 & 20.62  & 28.51  & 36.44 \\
Toys         & 61.19 & 23.25  & 30.54  & 41.75 \\
Grocery      & 66.22 & 24.77  & 32.44  & 44.07 \\
Clothing     & 63.63 & 25.14  & 33.30  & 42.58 \\
Software     & 85.71 & 46.23  & 55.95  & 67.66 \\ \hline
\end{tabular}
\end{table}

We also conduct experiments across different categories of MAVE.
Due to the limited space, Table~\ref{tab:category} reports the selected categories (the worst 5 and best 5 categories).
We observe that performance varies for different categories.
Some categories (i.e. software, grocery) can achieve better performance because the products in these categories have optical characters shown on the surface of the product and different products have distinct patterns.
Some categories (i.e. industrial, home kitchen, etc.) perform poorly because the patterns and features of the product images are quite similar and hard to distinct.
For future work, a category-oriented training process can be explored to train category-related text decoders separately.
\begin{table}[]
\caption{Ablation results over ViOC-AG components in the zero-shot setting on MAVE dataset.}
\small
\label{tab:ablation}
\centering
\tabcolsep=0.08cm
\begin{tabular}{lcccc}
\hline
              & 80\%Acc.  & Macro-F1 & Micro-F1 & ROUGE \\ \hline
w/o $D_{T}$      & 38.34 & 12.23  & 16.71  & 22.47 \\
w/o $M^{*}$   & 33.94      &   9.07     &   12.42     &  18.41     \\
w/o prompts      &  49.63     & 15.71       &  21.07      & 27.36      \\
w/o OCR       &  52.85     &   16.68     &     22.43   &  30.23     \\
ViOC-AG (All) &  \textbf{54.82}     &    \textbf{17.71}    & \textbf{23.69}       &   \textbf{31.92}    \\ \hline
\end{tabular}
\end{table}

\begin{figure*}[htp] 
\center{\includegraphics[height=8.4cm,width=\textwidth]{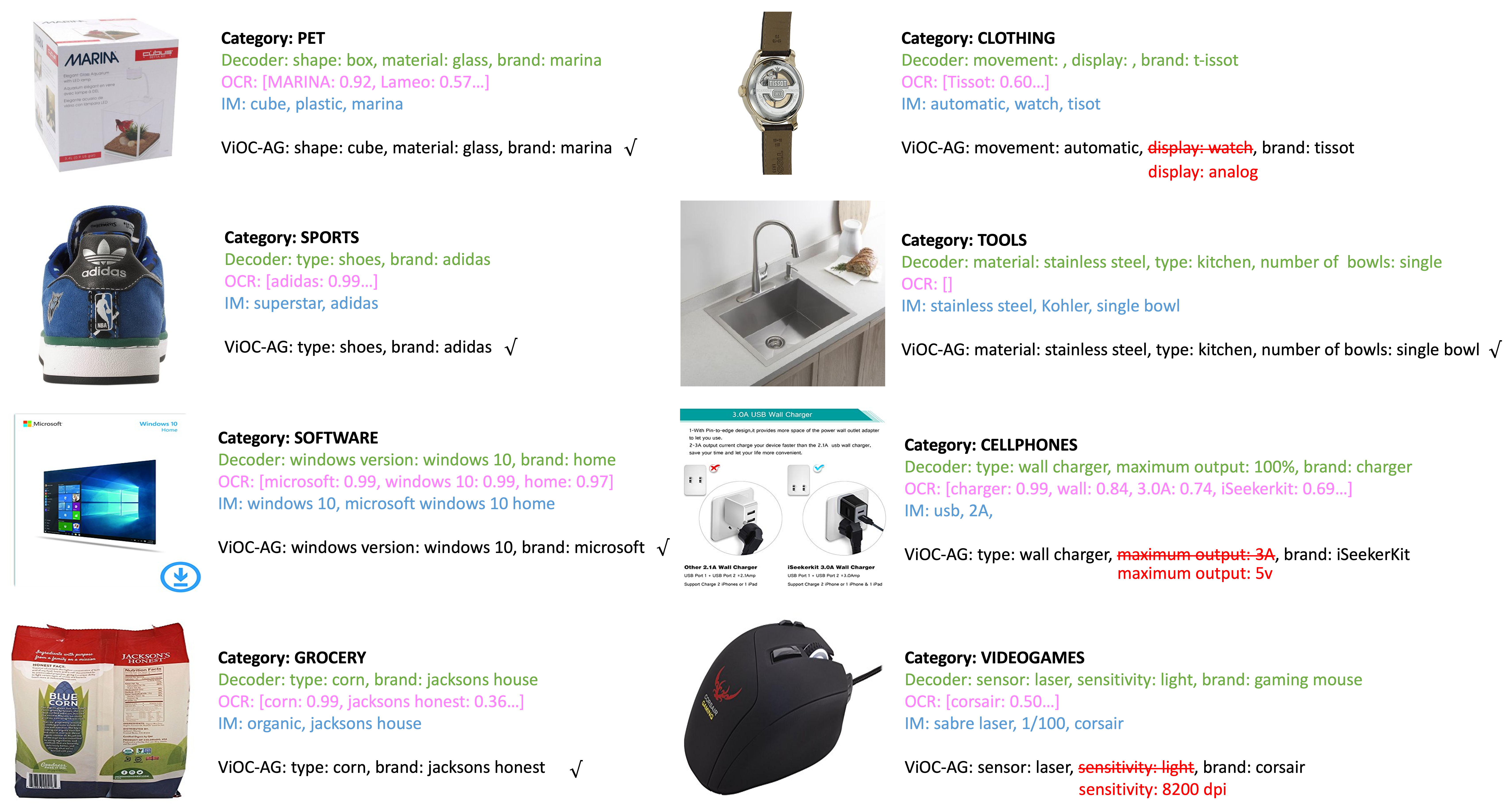}}
 \caption{\label{fig:case} Demonstrations of ViOC-AG for product attribute value generation across eight different categories.}
 \end{figure*}

 \begin{table*}[]
\caption{Results (\%) of 80\% Accuracy over ten attributes.}
\small
\label{tab:attribute}
\centering
\begin{tabular}{lcccccccccc}
\hline
               & Material       & Style          & \begin{tabular}[c]{@{}c@{}}Shoe \\ Style\end{tabular} & Form           & \begin{tabular}[c]{@{}c@{}}Clothing \\ Type\end{tabular} & Pattern        & Flavor         & \begin{tabular}[c]{@{}c@{}}Bowl \\ Shape\end{tabular} & Animal         & Color          \\ \hline
LLaVA          & 8.82           & 8.39           & 40.60                                                 & 20.86          & 37.71                                                    & 44.69          & 14.91          & 35.62                                                 & 29.27          & 16.67          \\
InstructBLIP   & 12.60          & 10.60          & 49.30                                                 & \textbf{27.20} & 50.01                                                    & 63.99          & 22.58          & 39.73                                                 & 35.56          & 35.90          \\
BLIP-2         & 13.88          & 10.80          & \textbf{77.40}                                        & 14.88          & 51.60                                                    & 61.94          & 23.53          & 42.47                                                 & \textbf{39.25} & 46.51          \\
ViOC-AG (ours) & \textbf{14.89} & \textbf{19.15} & 72.00                                                 & 15.96          & \textbf{52.14}                                           & \textbf{71.74} & \textbf{25.09} & \textbf{46.81}                                        & 39.22          & \textbf{50.00} \\ \hline
\end{tabular}
\end{table*}

\subsubsection{Ablation Study}
To verify the effectiveness of each part in ViOC-AG, we take ablation study in Table~\ref{tab:ablation}. 
We observe: 

(1) The task-customized decoder and the frozen LLM used in the training phase are important in ViOC-AG as the performance drops drastically when removing them.
We think it is because a pre-trained text decoder is usually used to generate long and diverse output descriptions. 
However, our task is quite different where the generated outputs are short phrases with specific formats.
There is no need for polishing the word but correcting the phrase in the generation process.
The outputs from the frozen LLM added to the original aspects inputs increase input data diversity, alleviating bias and overfitting for the trained text decoder.
(2) Fusing answers from the frozen prompt-based LLM and OCR systems to correct the final generated aspects is useful for ViOC-AG, which is consistent with our hypothesis that some attribute values (i.e. brand name, capacity, etc.) may appear on the product packaging.
To further improve the performance on out-of-domain aspect generation, a better customized OCR system, and diverse prompt templates can be explored in future work.

\subsubsection{Case Study}

\begin{table*}[]
\caption{Examples of aspects over ten different attributes.}
\small
\label{tab:example}
\centering
\begin{tabular}{l|l}
\hline
Attributes    & Aspects                                                                                          \\ \hline
Material      & {[}`leather', `wood', `stainless steel', `red rubber', `nylon', `canvas', `ceramic', `stoneware', `linen',...{]}                      \\
Style         &    {[}`casual', `knee high', `over-ear', `in-ear', `low-cut', `double-sided', `rotary', `brief', `everyday',...{]}                                                                                                 \\
Shoe Style    &  {[}`running shoe', `hiking boot', `walking', `skateboarding', `basketball', `golf', `soccer', `hunting',...{]}                                                                                                  \\
Form          &  {[}`whole', `crystal', `powder', `bag', `packet', `k-cup', `granular', `gel', `gallon', `spray paint',...{]}                                                                                                   \\
Clothing Type & {[}`sweater', `coat', `jacket', `hoodie', `raincoat', `shirt', `dress', `argyle', `jersey'{]}                                                                                                       \\
Pattern       &   {[}`plaid', `galaxy', `camo', `stripe', `polka dot', `flower', `camouflage', `argyle', `leopard', `solid',...{]}                                                                                                  \\
Flavor        &    {[}`buffalo', `vanilla', `chocolate', `lemon', `honey roasted', `chipotle', `sweet \& salty', `cinnamon',...{]}                                                                                                \\
Bowl Shape    & {[}`round', `elongated', `round-front'{]}                                                        \\
Animal        &  {[}`dog', `ferret', `cat', `puppy', `guinea pig', `rabbit', `hamster', `kitten', `canine', `chinchilla',...{]}                                                                                                \\
Color         & {[}`white', `manzanilla', `red', `rainbow', `chocolate', `blue', `green olives', `chardonnay', `pink'...{]} \\ \hline
\end{tabular}
\end{table*}

For the examples shown in Figure~\ref{fig:case}, the outputs from the task-customized decoder are shown in green. The OCR results are shown in pink and the outputs from the image caption model are shown in blue. 
Based on these examples, we observe that:

(1) In general, most of the attribute values can be generated from the trained task-customized text decoder. 
There are some cases in which the trained decoder may not generate correct attribute values. 
For example, in the videogames case, the decoder generates `gaming mouse' for the attribute of the brand.
We conjecture that this is probably because of the data distribution and features of the training data. There are limited data (product) samples with the attribute value of `brand: corsair' whereas there are lots of gaming mouse products in the training data.
This issue is solved by our correction stage using OCR characters and answers from the image caption model introduced in Sec.~\ref{sec:zero-shot}.
(2) OCR correction performs very differently among different categories. For the videogames case above, OCR can correct the brand name because `corsair' is shown on the mouse. However, characters seldom appear for some categories such as TOOLS. In such categories, OCR shows limited or even no performance improvement.
(3) In most cases, our proposed model ViOC-AG can correctly generate the attribute values after the correction stage for the trained text decoder. However, there still exists some difficult attributes such as `display', `maximum output', and `sensitivity'. 
These attributes are never directly shown as characters in the image. In addition, these attributes can be hardly learned from the visual features of the product image. 
Such difficult cases have the following features: (a) Attribute names are rare in the training set. For instance, `maximum output' and `sensitivity' may only be applied to some specific products; (b) The values include digital numbers. If the digital numbers are not shown directly in the image, our OCR module can not help to correct the attribute values. The numbers (i.e. 5v, 8200 dpi) can not be learned from the visual features.
These hard attributes need further exploration in future.


\subsubsection{Error Analysis}

To explore the attribute-level performance, we conduct experiments over ten randomly selected attributes reported in Table~\ref{tab:attribute}.
We observe that there is a significant variation in performance across different attributes among the models.
We conducted a more in-depth analysis of the dataset shown in Table~\ref{tab:example}.
For those showing better performance, for example,  different clothing types (hoodies v.s. dresses) can be differentiated by distince visual characteristics and design formats such as sleeve style, neckline, length, etc.

For those low-performance attributes, they have the following features: (1) The aspects can't be distinguished by visual features. For example, the flavor types (buffalo sauce v.s. honey roasted) are hard to be identified only by the image of the food as they may have similar color. The material (ceramic v.s. stoneware) is also challenging to be differentiated as they have manufacturing process overlaps (they both involve the firing of clay at high temperatures).
Combining image data with textual descriptions would be a potential solution. For example, the model can use textual descriptions or ingredient lists accompaning food images to infer flavor types.
(2) The aspects are very subjective. For example, two people are looking at the same food item, their interpretation of its flavor might differ based on personal taste and experience.
For the future work, confidence scores can be added for different interpretations, rather than deterministic outputs.
(3) The definitions for different aspects are quite vague, especially for terms like style and form. In these situations, the model is hard to learn and understand what exact information (aspects) the product image has. The model can be trained with in-context prompt learning on these aspect definitions and explanations to solve the ambiguous definitions in the future work.



\section{Conclusion}
In this paper, we formulate the attribute value extraction as a cross-modal generation task, which only requires product images as the inputs.
We propose ViOC-AG to generate unseen product aspects, which includes a text-only trainable projector and task-customized decoder to alleviate both the modality gap and task disconnection.
For zero-shot inference, ViOC-AG employs OCR tokens and results from a frozen prompt-based LLM to correct the decoded outputs for out-of-domain attribute values.
Results on MAVE demonstrate that our proposed model ViOC-AG outperforms other state-of-the-art fine-tuned vision-language models and it can achieve competitive results with textual generative LLMs, showing the bright future directions of cross-modal zero-shot attribute value generation.

\bibliography{custom}

\appendix

\section{Dataset Statistics}~\label{sec:data}
\begin{table}[h]
\caption{Dataset Statistics.}
\label{tab:dataset}
\centering
\begin{tabular}{lccc}
\hline
           & Train  & Validation & Test   \\ \hline
Products   & 403005 & 94426      & 188267 \\
Attributes & 620    & 560        & 576    \\
Aspects    & 44505  & 20148      & 33060  \\ \hline
\end{tabular}
\end{table}

\begin{figure}[!htb]
\centering
{
\includegraphics[width=3.7cm]{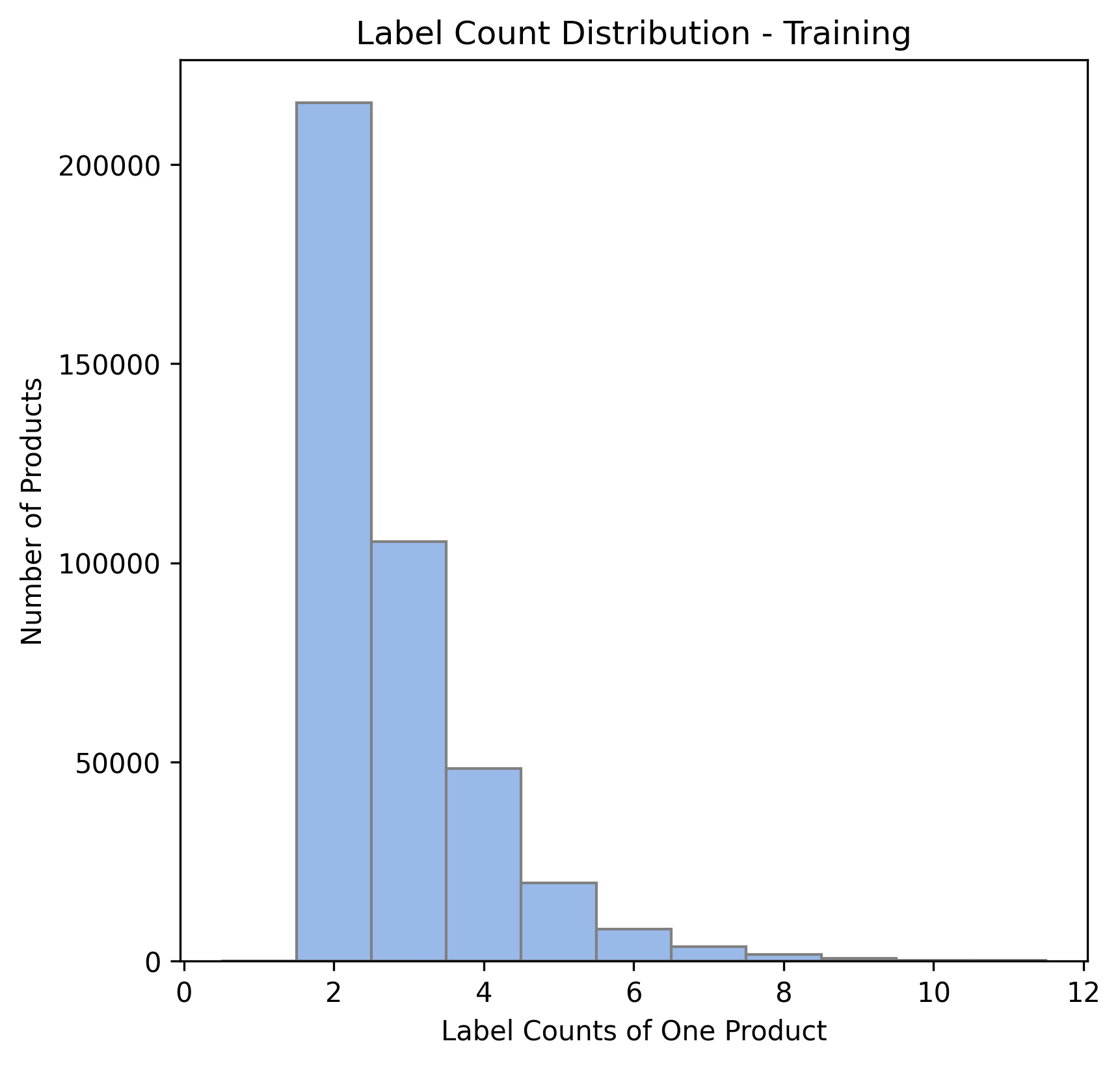}
}
\hspace{-4mm}
{
\includegraphics[width=3.7cm]{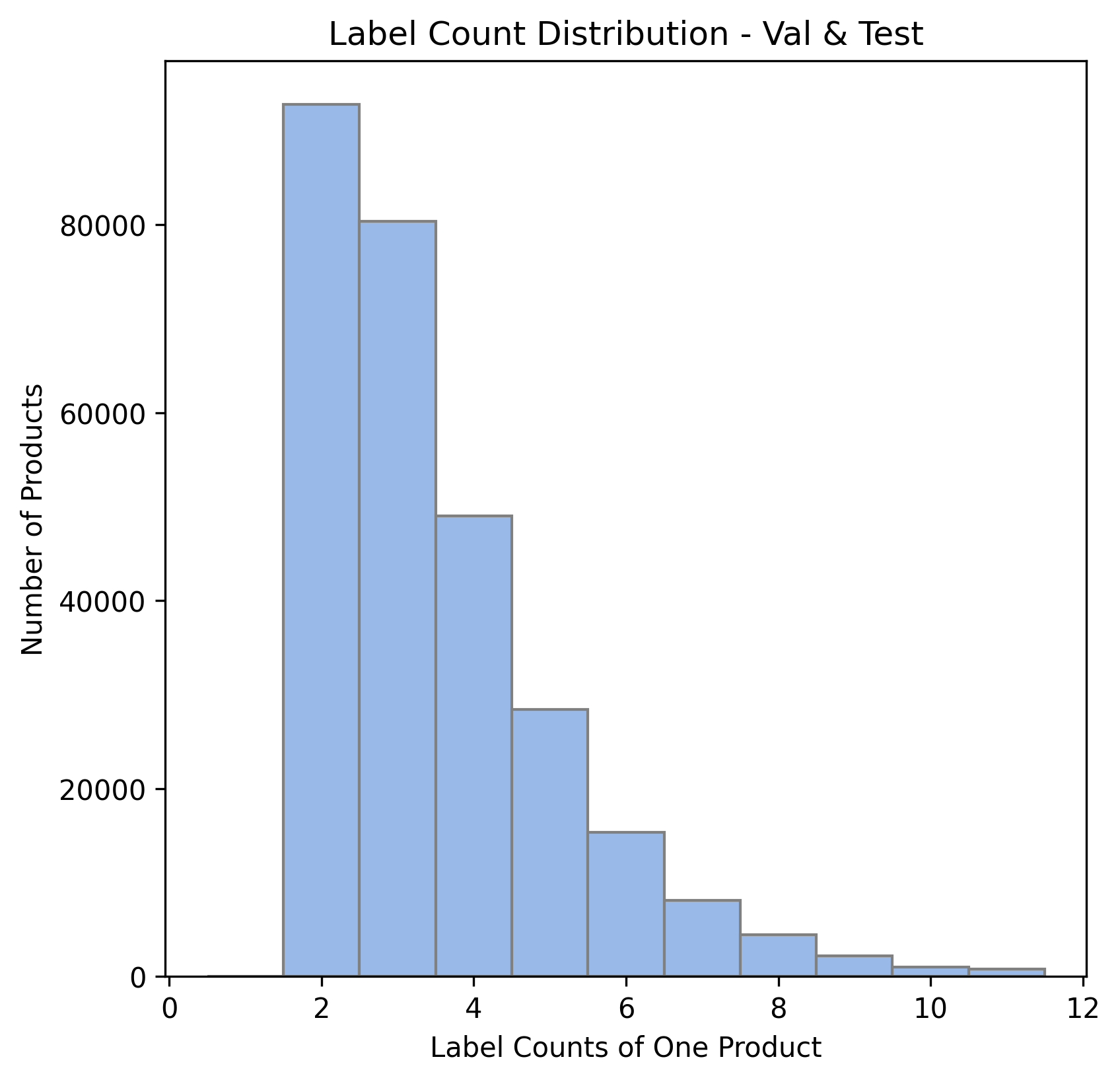}
}
\caption{Label Count Distribution.}
\label{fig:label} 
\end{figure}

The dataset statistics are shown in Table~\ref{tab:dataset}, where aspects are attribute values.
The distribution of label counts is shown in Figure~\ref{fig:label}.

\section{Evaluation Metrics}~\label{sec:evaluation}
We use 80\%Accuracy because the generative text decoder may generate more words than expected or generate words in the same meaning but with different forms (i.e. singular or plural forms), and we do not need a 100\% accuracy rate, which means all generated tokens are exactly correct with the ground truth.
For example, we consider the following aspects as the same aspect using 80\% Accuracy: `type: boot', `type: bootie' and `type: booty', `sleeve style: long sleeve', `sleeve style: long-sleeve' and `sleeve style: long sleeve length', etc.
We use F1-score because it is a balance of Precision and Recall. We follow~\cite{zou2024implicitave} to determine whether the generated answer is correct by checking whether the generated answer contains the true answer.
We use ROUGE as ROUGE focuses on recall, which means how much the words in the ground truth appear in the candidate model outputs.

\section{Parameter Setting}~\label{sec:parameter}
We randomly select unseen attribute value pairs following the sampling rule in Sec.~\ref{sec:data_sampling}.
For the hyperparameter and configuration of our proposed model ViOC-AG, we implemented ViOC-AG in PyTorch and optimized with AdamW optimizer. 
We train ViOC-AG and all baselines on the training set and we use a validation set to select the optimal hyper-parameter settings, and finally report the performance on the test set. We follow the early stopping strategy when selecting the model for testing. 
Our proposed model ViOC-AG achieves its best performance with the following setup. The learning rate is $0.0005$. The batch size is 512. The cosine similarity threshold $\tau_{d}$ is 0.95, the OCR token confidence $\tau_{c}$ is 0.5.
The experiments are conducted on eight Nvidia A100 GPUs with 80G GPU memory.

\end{document}